# Identifying Carbon as the Source of Visible Single Photon Emission from Hexagonal Boron Nitride

Noah Mendelson,[1] Dipankar Chugh,[2] Jeffrey R. Reimers,[1,3] Tin S. Cheng,[4] Andreas Gottscholl,[5] Hu Long,[6,7,8] Christopher J. Mellor,[4] Alex Zettl,[6,7,8] Vladimir Dyakonov,[5] Peter H. Beton,[4] Sergei V. Novikov,[4] Chennupati Jagadish,[2,10] Hark Hoe Tan,[2,10] Michael J. Ford,[1] Milos Toth,[1,10] Carlo Bradac,[1] Igor Aharonovich[1,10]*

[1]School of Mathematical and Physical Sciences, University of Technology Sydney, Ultimo, New South Wales 2007, Australia.
[2]Department of Electronic Materials Engineering, Research School of Physics and Engineering, The Australian National University, Canberra, Australian Capital Territory, Australia
[3]International Centre for Quantum and Molecular Structures and Department of Physics, Shanghai University, Shanghai 200444, China.
[4]School of Physics and Astronomy, University of Nottingham, Nottingham NG7 2RD, UK
[5]Experimental Physics 6 and Würzburg-Dresden Cluster of Excellence, Julius Maximilian University of Würzburg, Würzburg, Germany.
[6]Department of Physics, University of California, Berkeley, CA, USA.
[7]Materials Sciences Division, Lawrence Berkeley National Laboratory, Berkeley, CA, USA.
[8]Kavli Energy NanoSciences Institute at the University of California and the Lawrence Berkeley National Laboratory, Berkeley, CA, USA.
[9]ARC Centre of Excellence for Transformative Meta-Optical Systems, University of Technology Sydney, Ultimo, New South Wales, Australia.
[10]ARC Centre of Excellence for Transformative Meta-Optical Systems, Research School of Physics and Engineering, The Australian National University, Australian Capital Territory, Australia

*Igor.Aharonovich@uts.edu.au

**Abstract:**
Single photon emitters (SPEs) in hexagonal boron nitride (hBN) have garnered significant attention over the last few years due to their superior optical properties. However, despite the vast range of experimental results and theoretical calculations, the defect structure responsible for the observed emission has remained elusive. Here, by controlling the incorporation of impurities into hBN and by comparing various synthesis methods, we provide direct evidence that the visible SPEs are carbon related. Room temperature optically detected magnetic resonance (ODMR) is demonstrated on ensembles of these defects. We also perform ion implantation experiments and confirm that only carbon implantation creates SPEs in the visible spectral range. Computational analysis of hundreds of potential carbon-based defect transitions suggest that the emission results from the negatively charged $V_BC_N^-$ defect, which experiences long-range out-of-plane deformations and is environmentally sensitive. Our results resolve a long-standing debate about the origin of single emitters at the visible range in hBN and will be key to deterministic engineering of these defects for quantum photonic devices.

Single defects in solids have become some of the most promising frontrunner hardware constituents of applications in quantum information technologies and integrated quantum photonics.[1] Significant effort has been devoted to isolate and deterministically engineer such defects in wide bandgap materials such as diamond and silicon carbide.[2, 3] This collective effort resulted in spectacular proof of principle demonstrations ranging from quantum networks to spin-photon interfaces, [3] while simultaneously and steadily leading to understanding the fundamental level structures of these defects.

Recently, hexagonal boron nitride (hBN) has emerged as a promising host material for defects which display ultra-bright single photon emission (SPE)[4-8]. They exhibit remarkable properties: a strong response to applied strain and electric fields (Stark shifts),[9-11] stability under high pressure and elevated temperatures,[12, 13] potential for resonant excitation above cryogenic temperatures,[14-16] and addressability via spin-selective optical transitions.[17, 18] Yet, despite the numerous experimental characterizations and in-depth theoretical attempts to model their possible crystalline structure,[6, 19-23] the nature of these defects remains unknown.

Part of the challenge stems from standard hBN bulk crystal synthesis *via* high pressure high temperature not being amenable to the deterministic control of impurity incorporation. This is aggravated by the induced impurities often segregating and forming regions of inhomogeneous defect concentration.[24] In addition, the two-dimensional, layered nature of hBN makes ion implantation difficult to control. These limitations have precluded identifying the exact origin of the single photon emission in the material.

Here, we address this problem by carrying out a detailed study surveying various hBN samples grown in different laboratories by metal-organic vapor phase epitaxy (MOVPE) and molecular beam epitaxy (MBE). We find compelling evidence that to observe photoluminescence from SPEs the inclusion of carbon atoms in hBN is required. By systematically growing samples with different carbon concentrations, we show that the carbon content determines whether the photoluminescence signal originates from an ensemble of emitters (high carbon concentration) or isolated defects (low carbon concentration). Defect ensembles are demonstrated to display room temperature optically detected magnetic resonance (ODMR). We carry out multi-species ion-implantation experiments on MOVPE films and rule out the possibility that the emitters are associated to native vacancy complexes. Our results are supported by rigorous modelling analysis of carbon related defects.

Table 1 summarizes the materials analyzed. They are epitaxial hBN samples grown by different methods and under various conditions. The rationale was to understand whether the single defects are intrinsic to hBN (e.g. substitutional or interstitial nitrogen or boron complexes) or they involve foreign atoms (e.g. carbon). We investigated hBN samples grown by four methods. 1) Metal organic vapor phase epitaxy (MOVPE) with varying flow rates of the precursor triethyl boron (TEB)—a parameter known to systematically alter the levels of incorporated carbon. 2) Molecular beam epitaxy (MBE) on sapphire with and without a source of carbon. 3) High-temperature MBE on SiC with a varying orientation of the Si face to explore the possibility of carbon incorporation occurring from the substrate. 4) Growth by the conversion of highly oriented pyrolytic graphite (HOPG) into hBN. Note, that in the current work we focus on bottom up growth of hBN as it offers an opportunity for large (centimeter) scale films of desired thickness (down to ~1 nm), as well as better control over the inclusions of impurities. The table includes growth details, sample characteristics, shorthand names used throughout the rest of the paper, and general emission properties. Additional synthetic details are provided in the experimental section.

| Sample | Abbreviation | Growth Method & Details | SPE Photoluminescence | Additional Info |
|---|---|---|---|---|
| MOVPE hBN (TEB flux 10) | MOVPE hBN (TEB 10) | MOVPE on Sapphire, Precursors (triethyl borane & ammonia) TEB flow 10 µmol/min, H$_2$ carrier gas, 1350°C | Isolated SPEs, ZPLs Predominantly ~585±10 nm | ~40 nm thick |
| MOVPE hBN (TEB flux 20) | MOVPE hBN (TEB 20) | MOVPE on Sapphire, Precursors (triethyl borane & ammonia) TEB flow 20 µmol/min, H$_2$ carrier gas, 1350°C | Dense and Uniform Ensemble of SPEs with ZPL ~585, PSB ~630. | ~40 nm thick |
| MOVPE hBN (TEB flux 30) | MOVPE hBN (TEB 30) | MOVPE on Sapphire, Precursors (triethyl borane & ammonia) TEB flow 30 µmol/min, H$_2$ carrier gas, 1350°C | Dense and Uniform Ensemble of SPEs with ZPL ~585, PSB ~630. | ~40 nm thick |
| MOVPE hBN (TEB flux 60) | MOVPE hBN (TEB 60) | MOVPE on Sapphire, Precursors (triethyl borane & ammonia) TEB flow 60 µmol/min, H$_2$ carrier gas, 1350°C | Dense and Uniform Ensemble of SPEs with ZPL ~585, PSB ~630. | ~40 nm thick |
| MBE hBN on Sapphire | Undoped MBE hBN on sapphire | MBE on sapphire, Boron flux from e-beam source (300W). Boron in BN crucible. Nitrogen flow 2sccm. Growth temperature 1250°C. | No SPEs Present | ~20 nm thick |
| MBE hBN on Sapphire with Carbon Crucible | Carbon doped MBE hBN on Sapphire | MBE on sapphire, Boron flux from e-beam source (210W). Boron in BN crucible. Nitrogen flow 2sccm. Growth temperature 1250°C. | Semi-Isolated SPEs, ZPLs Range from 570-770 nm, Density ~5-8/µm$^2$ | ~18 nm thick |
| MBE hBN on Silicon Carbide (0° Si Face) | Undoped MBE hBN on SiC (0°) | SiC (Si-face, orientation-on) MBE on SiC. Boron flux from HT Knudsen source at 1875°C. Nitrogen flow 2sccm. Growth temperature 1390°C. | Very Few SPEs Density ~1 SPE in 40µm$^2$ | ~3 nm thick |
| MBE hBN on Silicon Carbide (8° Si Face) | Undoped MBE hBN on SiC (8°) | SiC (Si-face, orientation 8°-off) MBE on SiC. Boron flux from HT Knudsen source at 1875°C. Nitrogen flow 2sccm. Growth temperature 1390°C. | Isolated SPEs, ZPLs Range from 575-735 nm Density ~3-5/µm$^2$ | ~7 nm thick |
| HOPG→hBN Conversion | Converted hBN | HOPG is placed in a radio frequency induction furnace at 2000°C, N$_2$ gas is mixed with thermalized B$_2$O$_3$ powder facilitating conversion | Dense and Uniform Ensemble of SPEs with ZPL ~580, PSB ~630. | Bulk |

*Table 1—Epitaxial hBN Samples with Varying Carbon Concentrations. The 9 different hBN sample types used in the study, their growth methods, and SPE characteristics. Color coding correlates with the growth method.*

We first explore the photoluminescence (PL) from a series of hBN samples grown by MOVPE[25] as the triethyl boron (TEB) flow rate is increased and the ammonia flow is kept constant shown in Figure 1a. The aim of this measurement is to engineer an ensemble of hBN emitters, and to compare their properties with isolated SPEs grown using the same growth technique. A region of the TEB 10 (µmol/min) sample with the lowest percentage of carbon shows negligible fluorescence. Increasing the flow rate to TEB 20 is accompanied by the appearance of a bright fluorescence signal with two clear peaks appearing at ~585 nm and ~635 nm. Further increasing the flow rate to TEB 30 and 60 provides a similarly structured PL signature, with higher fluorescence intensity, confirming that higher PL intensity directly correlates with higher TEB flux. Moderate fluctuations in the peak positions, and intensity ratio of the 585 nm and 635 nm peaks at different sample locations are consistent with emission from dense ensembles of hBN emitters. This also confirms previous findings showing that hBN emitters possess zero-phonon line (ZPL) wavelengths clustered at ~585 nm when the sample is grown epitaxially.[26, 27] The energy detuning between the ZPL of the ensemble and phonon sideband (PSB) peak is ~176 meV on average.[28, 29] In specific examples, we can resolve the energy detuning between the ZPL and each of the two-known longitudinal-optical (LO) phonon modes, LO$_1$ ~165 meV and LO$_2$ ~195 meV,[28, 29] see supplementary information.

Previous studies on these samples have confirmed that with sufficiently slow growth kinetics (i.e. TEB 10) excess carbon atoms are largely removed in the ammonia atmosphere leaving a low percentage of carbon incorporation during growth.[25] With increasing precursor flux, layer-by-layer growth proceeds faster than the process of carbon removal—resulting in an increase of carbon incorporation in the hBN matrix. This is confirmed by X-ray photoelectron spectroscopy

(XPS), and subsequent analysis of peak intensities in the C1s region associated with C-B and C-N bonds, see supplementary information. Figure 1b(c) demonstrate near linear correlations between C-B (C-N) bonding and increasing TEB flux, with C-B bonding being roughly an order of magnitude more prevalent than C-N bonding. Preferential formation of C-B bonds follows logically from noting the B species are introduced with three pre-existing bonds to C. PL intensity of the resulting ensemble emission likewise displays a linear correlation with carbon concentration, see supporting information. Based on these results, we advance that the SPE emission at ~580 nm in hBN is likely to originate from a carbon-related defect complex.

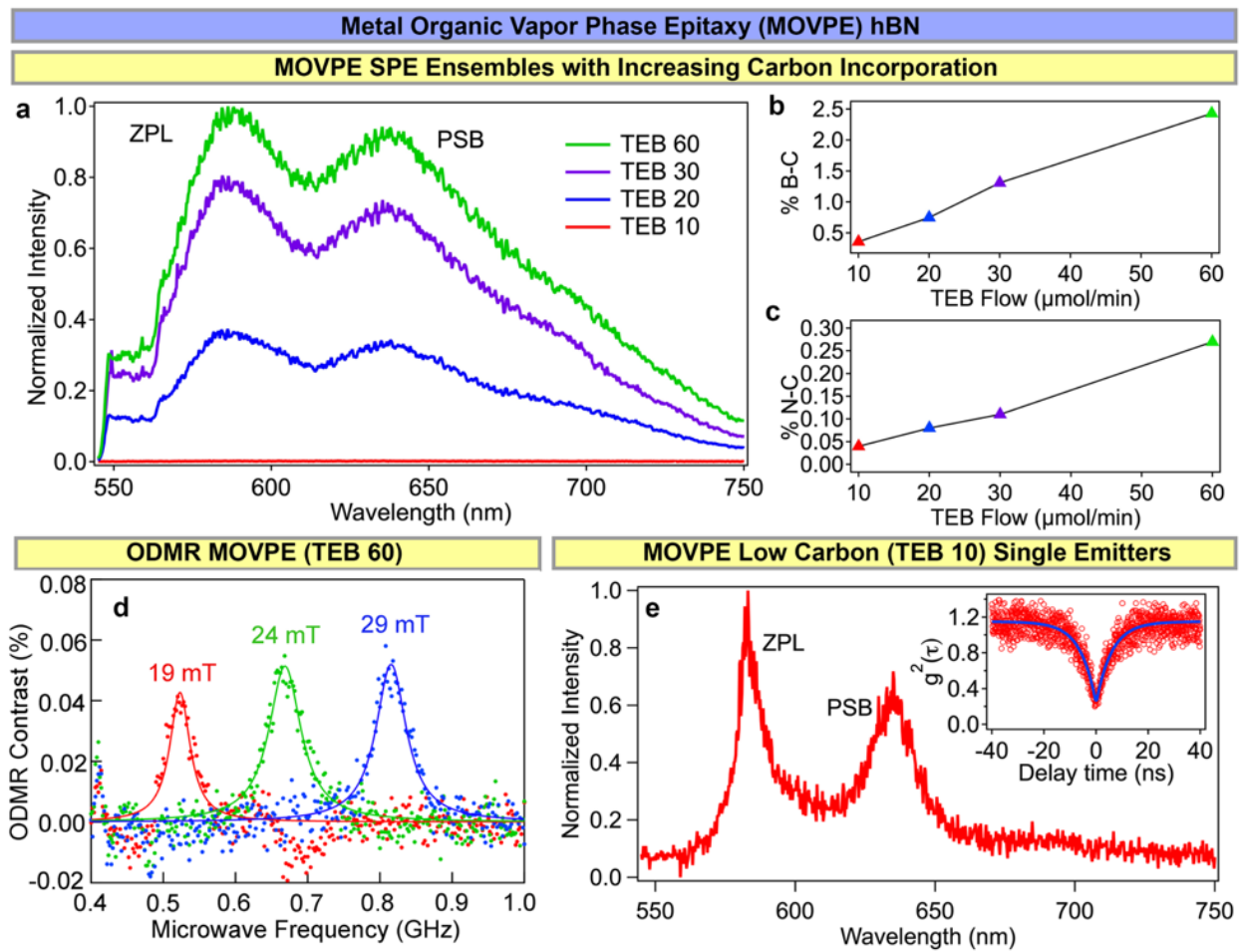

*Figure 1—Photoluminescence from MOVPE hBN Samples. **a.** MOVPE hBN grown with increasing flow rates of triethyl borane (TEB). As TEB flow increases, the fluorescence of SPE ensembles increases. **b.** Percentage of B-C bonding with increasing TEB flow evaluated by XPS. **c.** Percentage of N-C bonding with increasing TEB flow evaluated by XPS. **d.** Room temperature ODMR contrast observed from the ~585 nm ensemble emission of MOPVE hBN (TEB 60) at applied fields of 19, 24, and 29 mT respectively. **e.** Spectrum of a representative SPE found in MOVPE hBN TEB 10. Inset displays the corresponding autocorrelation measurements from the spectrum.*

Inspired by recent work showing that hBN defects emitting in the visible region can exhibit ODMR,[18] we performed such a measurement on ensembles of hBN emitters in MOVPE hBN (TEB 60). Figure 1d shows the ODMR spectra displayed as relative contrast, spin-dependent variation in photoluminescence ($\Delta PL/PL$), for different values of the static applied magnetic field. Our results are in line with some preliminary reports suggesting that this ODMR signal is potentially associated with a carbon-related defect.[18] However, importantly in previous experiments, ODMR was only observed at low temperature (8.5 K), while our measurements show that the ODMR contrast is resolvable at room temperature. By varying the static magnetic field B, we measure resonances at ~523, ~668.5 and ~815.4 MHz for B = 19, 24 and 29 mT, respectively. A value for $g_e$ of ~2.09 is extracted, as shown in the supporting information.

We next employ a lab-built confocal PL setup with a 532 nm excitation source, to study in detail the TEB 10 sample. The level of carbon doping is such that we can isolate single quantum emitters; a representative spectrum for one such emitter is shown in Figure 1d. The quantum nature of the emission was confirmed by measuring the second order auto-correlation function; the value of $g^{(2)}(\tau = 0) < 0.5$ (Fig. 1d inset) is conventionally attributed to a single photon source with sub-Poissonian emission statistics. We measured the zero-phonon line (ZPL) wavelength of 77 SPEs in the MOVPE hBN (TEB 10) sample, finding that ~78% of the emitters are located at (585 ± 10) nm, and 95% at wavelengths < 600 nm, see supporting information, consistent with previous studies on epitaxially grown hBN.[26, 27] The typical line shape of these emitters at room temperature is also consistent with previous studies, including the ZPL and a PSB centered at ~177 meV from the ZPL energy. This suggests that when the carbon concentration is sufficiently low, individual quantum emitters can be isolated. Their optical properties and spectral distribution are consistent with those observed in samples with higher carbon doping, with the difference merely being due to the density of emitters. This reinforces the hypothesis of a common—carbon-related—structural origin of these defects (see discussion below).

To further confirm that carbon-based defects are responsible for SPE emission from hBN we analyze a series of hBN samples grown by a different method, MBE[30]. Figure 2a displays the PL spectrum observed from undoped MBE hBN grown on sapphire substrate using an e-beam evaporator to supply boron from a boron nitride crucible. The resulting PL signal was relatively low; no SPEs could be found despite the material being of good quality as shown by a clear hBN $E_{2g}$ Raman line. Interestingly however, when the elemental boron precursor was placed inside a carbon crucible—with otherwise identical growth conditions—we observed the appearance of sharp spectral lines, shown in figure 2b. The carbon crucible used for e-beam thermalization of the boron source shows clear signs of sidewall etching, which suggests that carbon was present in the gas phase during growth.

In contrast to the undoped MBE hBN on sapphire showing only the hBN Raman (~573.7 nm) and the Cr in sapphire line (~694 nm), we find clear emission lines, corresponding to hBN SPEs. The SPEs are incorporated at a high density, up to 5 per confocal spot (~400 nm). Due to the high density of emitters incorporated, we could not unambiguously confirm their quantum nature via second-order autocorrelation measurements. We instead probed the polarization dependence of the emission by placing a polarizer in the collection path. Figure 2c shows one such collection, where emission from a presumed ZPL at ~577 nm is linearly polarized, with the PL intensity dropping when the polarizer is perpendicular to the polarization direction of the probed emitter (see more details in supporting information).

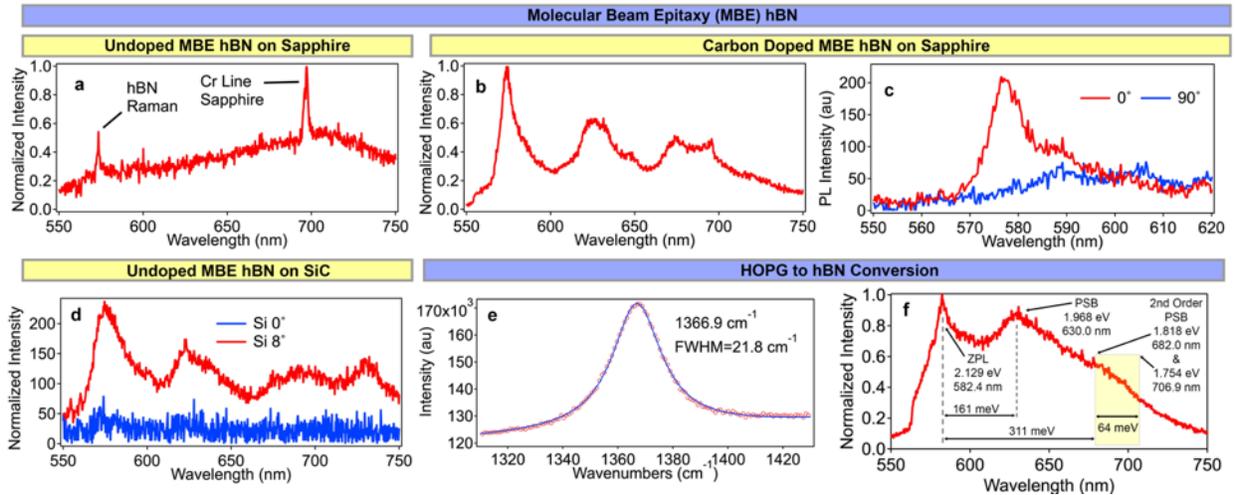

*Figure 2—Photoluminescence from MBE and HOPG Conversion hBN Samples.* ***a.*** *Undoped MBE hBN on sapphire displays no SPEs.* ***b.*** *Carbon doped MBE hBN on sapphire displaying a number of isolated peaks spanning the visible range. Typically, many SPEs are found within the laser excitation spot.* ***c.*** *Polarization resolved photoluminescence of a single peak in carbon doped MBE hBN on sapphire (b), demonstrating the polarized nature of the emission.* ***d.*** *Undoped MBE hBN on silicon carbide, with the Si face oriented at 0˚ (blue) and 8˚ off (red). While growth of the Si 0˚ face shows virtually no SPEs, growth on the Si 8˚-off face effectively incorporates SPEs via diffusion of C from the SiC substrate.* ***e.*** *Raman spectra of the HOPG to hBN conversion sample.* ***f.*** *Converted hBN displays an SPE ensemble emission centered around ~585 nm.*

We next explored MBE growth of hBN on silicon carbide (SiC), investigating different crystal orientations: specifically, with the top Si face-on (0˚) and slightly off (8˚). All hBN films grown on SiC utilized a Knudsen high temperature source for boron evaporation, at a growth temperature of 1390 ˚C, and identical conditions other than the substrates. Representative spectra from both sample types (Si at 0˚ and at 8˚) are displayed in Figure 2d. When growth was performed with the Si face at 0˚, only a single SPE peak was located across a 40 μm$^2$ scan. In contrast, when the Si face is oriented at 8˚ we found a number of SPE peaks across the sample, often with a number of different ZPL wavelengths appearing within the same confocal spot. The incorporated SPEs display a similar ZPL distribution to the carbon doped MBE hBN on sapphire (see supporting information), at a slightly lower density.

We attribute the incorporation of these SPEs during hBN growth on SiC to carbon diffusion from the substrate. At the growth temperature of 1390 ˚C, some sublimation of Si from the surface of the SiC substrates is expected, with the subsequent formation of an extra carbon layer on the surface of SiC.[31] While these temperatures are sufficient to sublime Si, they are not sufficient to evaporate C from the SiC surface.[31] Note that these temperatures are close to those used for graphene formation by SiC annealing,[31] suggesting that diffusion of carbon provides an available source for incorporation into the growing hBN layer. Interestingly, C incorporation into hBN appears significantly enhanced when the Si face is oriented 8˚ out of plane. The observed dependence of SPE incorporation during MBE growth further support the role of carbon in the origin of hBN SPEs in the visible spectral range.

Finally, we analyze a third technique for hBN growth, the conversion of HOPG to hBN, known to yield high quality porous hBN.[32] Conversion was confirmed by Raman spectroscopy (Figure 2e), with the hBN $E_{2g}$ mode located at 1366.9 cm$^{-1}$ with a FWHM of 21.8 cm$^{-1}$.[33] The conversion from graphite, proceeding via atomic substitutions, provides a high availability of carbon for incorporation as defects in the resulting hBN. Figure 2f displays a typical PL spectrum from the sample. We observe a bright and structured emission with ZPL and PSB peaks displaying similar transition energies as observed for high carbon MOVPE ensembles. Additionally, the resolved energy detuning of the peaks is similarly consistent with a ZPL and first and second order PSBs expected for hBN SPEs,[28, 29] leading us to ascribe the emission to an SPE ensemble. In certain confocal spots, we observed extremely bright SPEs; they undergo intermittent blinking events. Figure 2f displays such an example, while at other spots a broader emission is observed.

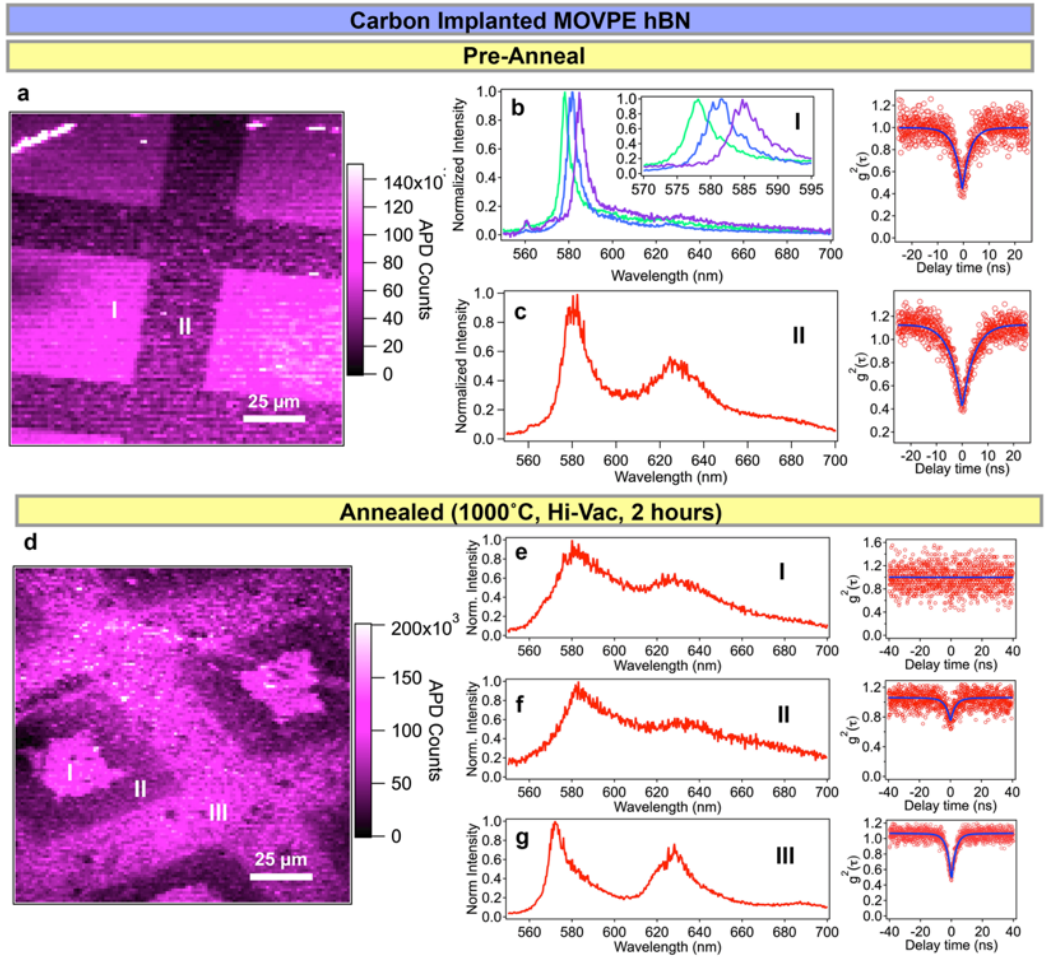

*Figure 3—MOVPE hBN (TEB 10) Samples Implanted with Carbon. Implantations were done at a dose of $10^{13}$ cm$^{-2}$ and an energy of 10 keV, using a TEM grid with 50 μm$^2$ apertures as a mask. **a**. Confocal scan of carbon implanted sample, where square areas marked (I) were implanted, and those marked (II) were masked. **b**. Spectra from implanted areas (I) display narrow ZPLs with almost no PSB and are attributed to emitters created via implantation. A representative $g^{(2)}(\tau)$ is shown to the right. **c**. Spectra and $g^{(2)}(\tau)$ from the masked area (II) display borderer ZPLs, and prominent PSBs. **d**. Confocal scan of the carbon implanted sample, post annealing, where areas marked (I) and (II) were implanted, while area (III) was masked. **e**. A representative spectra and $g^{(2)}(\tau)$ from area (I), showing an ensemble of hBN emitters, and a corresponding $g^{(2)}(\tau)$ measurement showing no dip as expected for ensemble emission. **f**. A representative*

*spectra and $g^{(2)}(\tau)$ from area (II) showing evidence of quantum emission but with significant spectral contributions from nearby SPEs resulting in a $g^{(2)}(\tau)$ value of ~ 0.75. **g.** A representative spectra and $g^{(2)}(\tau)$ from the masked area (III) post annealing showing a well-resolved SPE and PSB and a $g^{(2)}(\tau)$ confirming a single emission center.*

We now turn our attention to using ion implantation for defect creation, in an attempt to confirm the role of carbon. We performed a series of implantation experiments (dose: $10^{13}$ ions/cm$^{-2}$, energy 10 keV) with carbon as well as silicon and oxygen used as controls to rule out the possibility for the photoemission to be due to native vacancy defects. The implantation experiments were performed on the MOVPE hBN (TEB 10) films (film thickness ~ 40 nm, see supporting information) so to compare the relevant results to those for the samples synthesized while increasing carbon content during growth.

Figure 3a shows the confocal scan of an MOVPE hBN (TEB 10) film after carbon implantation, but prior to annealing, where a TEM grid with 50 µm$^2$ square apertures was used as a mask. The implanted region is labelled I, while the masked region is labelled II. Figure 3b displays spectra collected from emitters within the implanted region (I), and a representative $g^{(2)}(\tau = 0) < 0.5$, confirming the quantum nature of the emission from these centers. Figure 3c displays a representative emitter from the masked region (II), showing the typical line shape of the ZPL and the PSB peaks found in MOVPE hBN (TEB 10) films, with the corresponding $g^{(2)}(\tau = 0)$ shown to the right.

Inside the C-implanted region, most emitters (~ 80%) display narrow ZPL peaks (~5 nm FWHM) and extremely weak PSBs, compared to the typical ZPL/PSB found in these MOVPE hBN (TEB 10) films. For comparison, among 77 SPEs investigated in pristine MOVPE hBN (TEB 10) films the narrowest linewidth was ~8 nm, demonstrating the unique nature of these emission lines, see supporting information. The remaining ~20% of SPEs within the implanted region display similar line shapes (typically ~20 nm FWHM) and phonon coupling to those for the emitter in figure 3c and are attributed to preexisting SPEs in the region. Our results indicate that the sharp emission lines belong to SPEs created via implantation of C ions. The reasons for the observed narrow line shape and the minimal phonon coupling are explored further *via* computational modelling below.

After characterizing the implanted hBN films, the samples were then annealed in high vacuum (1000 ˚C, <10$^{-6}$ Torr, 2 hours), and the same set of measurements was performed. As shown in figure 3d, the implanted regions are still visible, they however show variations in PL intensity. This effect is likely due to ion scattering around the mask edges and vacancy diffusion—which have been observed for implantation in diamond.[34] The PL spectra from three different areas are shown in figure 3e–f, and correspond to (I) the implanted region of high PL intensity, (II) the implanted region of lower PL intensity, and (III) the masked region of the film.

Figure 3e displays a representative spectrum from inside region I, where we found emission characterized by broad ZPL and PSB profiles similar to those observed in the high TEB flux growths. This emission is confirmed to be due to an ensemble of SPEs as the corresponding $g^{(2)}(\tau)$ measurement shows no anti-bunching despite the associated ZPL/PSB structure. A similar spectral signature is observed consistently throughout region (I), again implying the creation of an ensemble of C-based SPEs. Figure 3f displays a representative spectrum from the implanted region II, where we again observe luminescence with a similar line shape. The overall ensemble signal remains homogeneous in this region, although appears less dense and bright, and a $g^{(2)}(\tau)$ function shows a value of ~0.75, confirming the presence of fewer emitters within a confocal spot. Note that in both implanted areas (I and II) we no longer find the narrow emission lines with low phonon coupling found prior to annealing. Finally, figure 3g displays a representative spectrum from region III (masked area), showing a typical ZPL and PSB profile and a $g^{(2)}(\tau) < 0.5$, which confirms the quantum nature of the emission. As expected, this is similar to the pattern observed for the unimplanted samples. Control experiments implanting silicon and oxygen with otherwise identical conditions were also performed, but the emitters, either singles or ensembles were not observed (see supporting information for details).

In light of the implantation results, we briefly consider a potential ancillary role of carbon. This could potentially occur through the stabilization or charge state modification of alternative defects, as well as modification of the material fermi level. Critically, our implantation results allow us to rule out these possibilities. The creation of SPEs prior to annealing with C implantation only (i.e. not with Si and O implantation), despite clear evidence of increased vacancy creation, precludes the potential of a secondary role of carbon in *activating* simple native vacancy complexes. Furthermore, complex native vacancies such as those involving substitution defects like $V_N N_B$, or non-carbon heteroatom impurities involving O and Si are similarly inconsistent with our results finding that only carbon implantation creates SPEs prior to annealing. Finally, the observation that high temperature annealing of the implanted samples leads to a spatially correlated and homogeneous ensemble emission only in the case of carbon implantation is further indicative of the structural inclusion of carbon in the emissive defect. The compiled evidence indicates in a compelling manner that the structural nature of hBN SPEs emitting in the visible is directly associated with carbon-based defect(s). We note the inclusion of carbon in even the highest purity hBN materials,[24] can explain the activation of hBN SPEs such as annealing, plasma treatment, and e-beam irradiation when carbon is not intentionally introduced,[35, 36] as is discussed further in the supplementary information.

To gain further insight into the structure of the carbon defect, we searched for defect transitions from which the observed photoemission could originate. To do so, time-dependent density-functional theory[37] (TD-DFT) calculations were performed using the CAM-B3LYP[38] density functional, see Supporting Information for details. Four main defect candidates were considered: $C_B$, $C_N$, $V_N C_B$, and $V_B C_N$ (Figure 4a-d) in their neutral, negative (–1), and positive (+1) charged states. Two spin manifolds were considered for each (either singlet and triplet or else doublet and quartet), as well as at least ten excited states of each type. Calculations were performed using 3-ring, 5-ring, and 10-ring model compounds containing 1 or 3 hBN layers, see supporting information. Figure 4e displays the 10-ring 3-layer model for $V_B C_N$-. Calculations on 10-ring systems were performed using a mixed quantum-mechanics / molecular-mechanics (QM/MM) scheme, utilizing an AMBER[39] potential fitted to mimic CAM-B3LYP results on 5-ring h-BN.

To eliminate unsuitable defect candidates, we focused on three well established experimental features of the SPEs, a ZPL energy transition of ~ 2.1 eV,[26, 27] a fast excited state lifetime of ~2-6 ns,[7] and a high quantum efficiency. Accordingly, computational results were filtered to reproduce first, a lowest-energy transition within each manifold of 1.6 – 2.6 eV (based on the expected *worst-case* computational error, calibrated for this method to be ±0.5 eV)[22], and second, an oscillator strength exceeding 0.1, compatible with the observed short photoluminescence lifetime and high quantum yield. As most transitions are predicted to have oscillator strengths one hundredth of this or much less, only two candidates appear to be of interest: the $(1)^4 B_1 \rightarrow (1)^4 A_2$ transition in $V_B C_N^-$, and the $(2)^3 B_1 \rightarrow (1)^3 B_1$ transition in $V_N C_B$. Furthermore, the later transition can be eliminated, as the transition is predicted to be broad and the ground-state of $V_N C_B$ is predicted to be $(1)^1 A'$. Hence only one viable defect possibility is identified, $V_B C_N^-$. Note that $C_N$, previously proposed as a candidate for the observed ODMR contrast,[18] was discarded due to its low oscillator strength (and thus long lifetime), one of the most reliable features of the calculation. Further details on $C_N$, $V_NC_B$, and all others simulated transitions are included in the supplementary information.

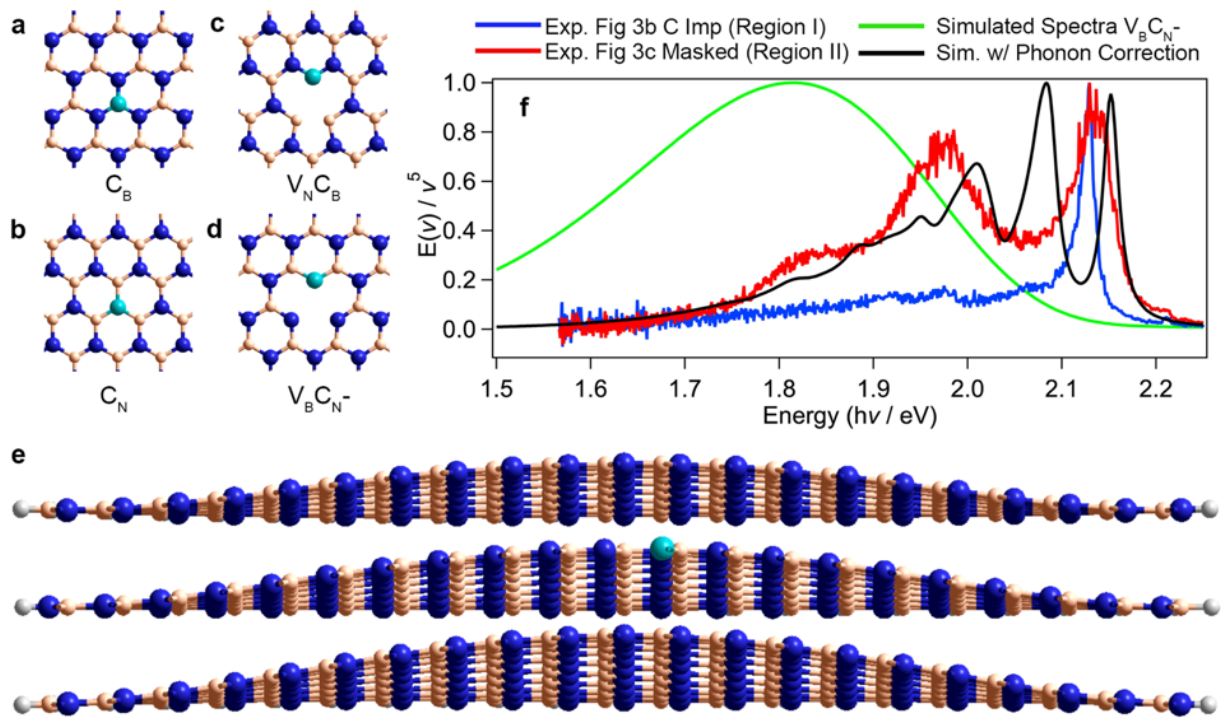

*Figure 4. Computational Modelling.* *Properties were determined for the $C_B$ (a), $C_B^+$, $C_B^-$, $C_N$ (b), $C_N^+$, $C_N^-$, $V_NC_B$, (c), $V_NC_B^-$, $V_NC_B^+$, $V_BC_N$, $V_BC_N^-$ (d-e), and $V_BC_N^+$ defects using model compounds up to 10 rings (5.1 nm) in diameter; N- blue, B- peach, C- cyan. It is concluded that only the $(1)^4A_2 \rightarrow (1)^4B_1$ transition in $V_BC_N^-$ could possibly account for the observed emission (f): red- Region-I spectrum ($\lambda^E = 0.13$ eV), blue- Region-II spectrum from Figure 1d ($\lambda^E = 0.15$ eV), green- predicted spectrum without correction for acoustic phonons, black- calculated spectrum treating all phonons below 0.06 eV as acoustic. The defect is predicted to distort its surrounding layers into bowl-like structures (see cross-section e from a 3-layer model); this could account for the observed large change in spectral bandshape at constant reorganization energy observed between Region-I and Region-II.*

The ground state of $V_BC_N^-$, is predicted to be $(1)^4A_2$, with unpaired electrons in $a_1$ ($\sigma$), $b_1$ ($\pi$) and $b_2$ ($\sigma$) orbitals. Its $(1)^4B_1 \rightarrow (1)^4A_2$ transition is of dominant $a_1(\sigma) \rightarrow b_2(\sigma)$ character, polarised in-plane and perpendicular to the defect's $C_{2v}$ axis, with an oscillator strength of 0.2. Large bowl-like out-of-plane deformations are predicted for both the ground and excited states. Such deformations are very sensitive to the environment, providing a plausible explanation for the large differences in spectral shape found in Region-I (Figure 3b) and Region-II (Figure 3c). Furthermore, arising from the bowl-like defect shape, spectra are expected to be especially sensitive to the location of the defect near step edges, surfaces, and other irregularities, as well as to strain and Stark effects. Spectra from both regions are presented in Figure 4f as the intrinsic defect spectral bandshape profile $E(\nu)/\nu^5$ (obtained as the raw emission scaled by wavelength to the 5th power). Despite very different bandshapes, both have similar emission reorganization energies $\lambda^E$ (average emission energies relative to the ZPL), consistent with a common origin of the emission. Shown also in the figure are simulated bandshapes before and after an applied correction for modelling acoustic phonons, the latter being consistent with the observed variations. Critically, the assignment of $V_BC_N$- as the likely origin of visible SPEs in hBN, is in line with the experimental findings herein. Additionally, the predicted deformations offer a plausible explanation for the diverse range of observed photophysical properties,[28, 40] and large tuning ranges demonstrated experimentally,[9] as is discussed further in the supporting information.

In summary, we have presented rigorous experimental results to confirm the central role of carbon in hBN quantum emitters in the visible spectral range. We compared samples grown by three different methods: MOVPE, MBE, and HOPG conversion. They all exhibited a direct correlation between the introduction of carbon as a precursor/substance and the formation of SPEs. Furthermore, MOVPE growth enabled us to deterministically control carbon incorporation and vary the density of the quantum emitters from singles to ensembles and observe room temperature ODMR. We have successfully reproduced equivalent results using direct ion implantation of carbon, confirming a carbon defect structure is responsible for visible spectrum quantum emission. Employing a TD-DFT method, we proposed the negatively charged $V_BC_N$- as the most suitable transition to explain the observed results and matched the calculated spectra to the observed ones experimentally. The predicted bowl-shaped distortions for this defect, making it especially susceptible to environmental influences, in line with the diverse range of emission energies and properties observed in visible region hBN SPEs. Our results will accelerate the deployment of visible quantum emitters in hBN into quantum photonic devices and will advance potential strategies for the controlled engineering of quantum emitters in van der Waals crystals.

**Acknowledgements**
The authors thank Lee Bassett and Audrius Alkauskas for fruitful discussions. This work at Nottingham was supported by the Engineering and Physical Sciences Research Council [grant number EP/K040243/1, EP/P019080/1]. We also thank the University of Nottingham Propulsion Futures Beacon for funding towards this research. We also acknowledge financial support from the Australian Research Council (via DP180100077, DE180100810 and DP190101058). Access to the epitaxial growth facilities is made possible through the Australian National Fabrication Facility, ACT Node. Ion implantation was performed at The Australian Facility for Advanced Ion Implantation Research (AFAiiR), RSP (ANU). This work was supported in part by the U.S. Department of Energy, Office of Science, Office of Basic Energy Sciences, Materials Sciences and Engineering Division under Contract No.DE-AC02-05-CH11231, within the sp2-Bonded Materials Program (KC2207), which provided for synthesis and structural characterization of h-BN converted from carbon.

**Methods**
*Metal Organic Vapor Phase Epitaxy.* hBN layers were grown on commercially available 2" sapphire substrates using metal organic vapor phase epitaxy (MOVPE), as described in.[25] Triethyl boron (TEB) and ammonia were used as the boron and nitrogen precursors, respectively, while hydrogen was the carrier gas. The precursors were introduced into the reactor as short alternating pulses, in order to minimize parasitic reactions between TEB and ammonia. hBN growth was carried out at a reduced pressure of 85 mBar and the growth temperature was set to 1350°C. In the present study, the TEB flux was varied from 10 µmol/min to 60 µmol/min to study the effect on carbon incorporation on sub-bandgap luminescence from the hBN films. For ion implantation, PL and SPE measurements, cm-sized hBN films were transferred from sapphire on to $SiO_2$/Si substrates, using water-assisted self-delamination.[25] Thickness of the hBN films was also measured using atomic force microscopy, as shown in the supplementary information. X-ray photoelectron spectroscopy was used for determining the impurity levels in the as-grown MOVPE-hBN films, as shown in the supplementary information. A gentle etching using Ar beam was performed in-situ to remove adventitious carbon and impurities from the surface; all spectra were collected from the bulk of hBN films.

*Molecular beam epitaxy.* BN epilayers were grown using a custom-designed Veeco GENxplor MBE system capable of achieving growth temperatures as high as 1850 °C under ultra-high vacuum conditions, on rotating substrates with diameters of up to 3 inches. Details of the MBE growth have been previously published.[41] In all our studies, we relied on thermocouple readings to measure the growth temperature of the substrate. For all samples discussed in the current paper the growth temperature was in the range 1250–1390 °C. We used a high-temperature Knudsen effusion cell (Veeco) or electron beam evaporator (Dr. Eberl MBE-Komponenten GmbH) for evaporation of boron. High-purity (5 N) elemental boron contains the natural mixture of $^{11}B$ and $^{10}B$ isotopes. To have boron in the e-beam evaporator we used boron nitride and vitreous carbon crucibles. We used a standard Veeco RF plasma source to provide the active nitrogen flux. The hBN epilayers were grown using a fixed RF power of 550 W and a nitrogen ($N_2$) flow rate of 2 sccm. We used 10 × 10 $mm^2$ (0001) sapphire and on- and 8°-off oriented Si-face SiC substrates.

*HOPG to hBN Conversion.* The conversion takes place in a graphite crucible. A HOPG crystal is placed in the center of the crucible on a separate graphite holder. Small holes in the stage holding the HOPG allow vapors from the boron-oxide powder, placed at the bottom of the crucible, to flow to the HOPG crystal. A radio frequency induction furnace is then heated to 2000 °C, and $N_2$ gas is introduced as the nitrogen precursor. A central tube mixing the nitrogen gas with the boron-oxide vapor pre-mixes the precursors prior to arriving at the HOPG crystal. Further details can be found here.[32]

*Ion Implantation.* Ion implantation was carried out on 40 nm-thick MOVPE-hBN films, grown using a TEB flux of 10 μmol/min. For this, the hBN films were first transferred on to $SiO_2$/Si substrates. A copper grid with a square mesh (GCu300, ProSciTech) was used as the implantation mask. Carbon, silicon and oxygen were separately implanted into the hBN films. During implantation, the ion energy and fluence were 10 keV and $10^{13}$ ion/$cm^2$, respectively.

*Confocal Microscopy.* PL studies were carried out using a lab-built scanning confocal microscope with continuous wave (CW) 532-nm laser (Gem 532, Laser Quantum Ltd.) as the excitation source. The laser was directed through a 532 nm line filter and a half-waveplate and focused onto the sample using a high numerical aperture (100×, NA = 0.9, Nikon) objective lens. Scanning was performed using an X−Y piezo fast steering mirror (FSM-300). The collected light was filtered using a 532-nm dichroic mirror (532 nm laser BrightLine, Semrock) and an additional 568-nm long pass filter (Semrock). The signal was then coupled into a graded-index multimode fiber (fiber aperture of 62.5 μm). A flipping mirror was used to direct the emission to a spectrometer (Acton Spectra Pro, Princeton Instrument Inc.) or to two avalanche photodiodes (Excelitas Technologies) in a Hanbury Brown-Twiss configuration, for spectroscopy and photon counting measurements, respectively. Correlation measurements were carried out using a time- correlated single photon counting module (PicoHarp 300, PicoQuant). All of the second-order autocorrelation $g^{(2)}(\tau)$ measurements were analyzed and fitted without background correction unless specified otherwise.

*ODMR.* The ODMR spectra were measured with a confocal microscope setup. A 100× objective (Olympus MPLN100X) was used to focus a 532-nm laser (LaserQuantum opus 532) onto the sample and collect the PL signal. The PL signal is collected back through a 650-nm short-pass dichroic mirror for separation from scattered laser light. Additionally, a 532nm and 550nm long-pass filter were used before the PL was detected by a silicon avalanche photodiode (Thorlabs APD440A) to filter out the laser light. The microwave field was applied through a signal generator-plus-amplifier system (Stanford Research Systems SG384 + VectaWave VBA1000-18 Amplifier); the sample was placed on a 0.5-mm-wide copper stripline. In order to detect the ODMR signal (i.e. the relative ΔPL/PL contrast) by lock-in technique (Signal Recovery 7230), the microwaves were driven with an on-off modulation. The resonant condition was changed with the external magnetic field by mounting a permanent magnet below the sample.

*Raman Spectroscopy.* Raman spectroscopy measurements were carried out on an In-Via confocal Raman (Renishaw) system using a 633-nm excitation source. Calibration of the Spectrometer was carried out using a Si substrate to 520 $cm^{-1}$. The peaks were then fitted to a Lorentzian line profile, from which the corresponding peak center position and full width at half maximum (FWHM) were extracted. Samples were analyzed after transfer to $SiO_2$.

*Computational.* Many innovative approaches are used in order to model defects with large 3D spatial deformations, as described in detail in Supporting Information. The core elements are the use of Gaussian-16 to perform TD-DFT calculations within a QM/MM model utilizing CAM-B3LYP and an AMBER h-BN force field fitted to mimic CAM-B3LYP. Spectra are simulated with the Huang-Rhys model based on analytically obtained second-derivatives for both ground and excited states.